\documentclass{pasa}%

\usepackage{graphicx}

\title[]{Can an Infinitely Long Object Fit in an Expanding Universe?}

\author{Aaron Glanville$^A$ $^*$, Tamara M.~Davis$^A$
 \affil{$^A$The University of Queensland, School of Mathematics and Physics, QLD 4072, Australia}
 \affil{$^*$E-mail: a.glanville@uq.edu.au}
}

\jid{PASA}
\doi{10.1017/pas.\the\year.xxx}
\jyear{\the\year}

\newcommand{\vt}{v_{\rm t}}
\newcommand{\vrr}{v_{\rm r}}
\newcommand{\vp}{v_{\rm p}}
\newcommand{\vtz}{v_{{\rm t},0}} 
\newcommand{\vrz}{v_{{\rm r},0}}
\newcommand{\vpz}{v_{{\rm p},0}}

\usepackage{aas_macros}
\usepackage{hyperref} 
\hypersetup{colorlinks,citecolor=blue,linkcolor=blue,urlcolor=blue}

\usepackage{soul}

\hypersetup{draft}

\begin{document}

\begin{frontmatter}
\maketitle

\begin{abstract}
Does space stretch its contents as the universe expands? Usually we say the answer is no --- the stretching of space is not like the stretching of a rubber sheet that might drag things with it. In this paper we explore a potential counter example --- namely we show that is is impossible to make an arbitrarily long object in an expanding universe, because it is impossible to hold the distant end of the object `stationary' with respect to us (as defined in the Friedmann-Lema\^{i}tre-Robertson-Walker metric). We show that this does not mean that expanding space has a force associated with it, rather, some fictitious forces arise due to our choice of reference frame. By choosing our usual time-slice (where all comoving observers agree on the age of the universe), we choose a global frame that does not correspond to the frame of any inertial observer. As a result, simple relativistic velocity transforms generate an apparent acceleration, even where no force exists. This effect is similar to the fictitious forces that arise in describing objects in rotating reference frames, as in the case of the Coriolis effect.
\end{abstract}

\end{frontmatter}

\section{Introduction}
\label{sec:level1}
The expansion of the universe is the cornerstone of modern cosmology \citep{Lemaitre1927,Hubble1929}. However, the description of cosmological recession in the framework of expanding space, and indeed the interpretation of expanding space itself, has been subject to significant debate in the literature \citep{Davis2003, Whiting2004, Barnes2006, GronElgaroy2007, Chodorowski2007, Francis2007, Peacock2008, Cook2009,  MacLaurin2015}. Such discussion has centred around two fundamental questions: ``What does it physically mean to say space is expanding?'' and ``Does this interpretation of expanding space provide a unique description of the observed cosmological recession?'' 

Through this paper, we focus on illuminating some of the unique features of expanding space by asking: ``Is there a limit to the size of an object that can be constructed in an expanding universe?'' This question is interesting, as in non-expanding space there is nothing that prevents us from making an arbitrarily large object. Even in the case of expanding space, we usually emphasise the ``stretching'' of space does not stretch or rip its contents.  Indeed, even separating space from its contents goes against the general relativistic perspective that ``Space tells matter how to move; matter tells space how to curve.'' \citep{MTW1973}.  However, in the standard Friedmann-Lema\^{i}tre-Robertson-Walker (FLRW) metric, space can expand faster than the speed of light \citep{Davis2004}. The Hubble-Lema\^{i}tre law holds, allowing us to define a distance known as the Hubble radius ($D_H = c/H$) beyond which cosmological recession is superluminal. Objects beyond this radius would require inward peculiar velocities ($v_{\rm p}$) faster than the speed of light to fix their position with respect to us. Since superluminal {\em peculiar} velocities are not possible, it would seem that something is indeed stopping us from making a non-expanding object larger than the Hubble sphere. In the absence of any force or tension, what feature of expanding space prevents us from constructing an arbitrarily large object?

This paper continues the tradition of \citet{Davis2003}, who explored the physical consequences of expanding space using the radial dynamics of test particles released in expanding universes. Their paper motivated the use of the tethered galaxy system as a thought experiment to explore what it means to say ``space is expanding''. Through this thought experiment, a massless test galaxy is defined with a fixed proper distance from some observer and zero proper velocity, analogous to a galaxy that is physically tethered to some central observer. This galaxy is then ``released'' and allowed to move freely. Using this system \citet{Davis2003} showed that in the limit of non-relativistic velocities, any induced radial motion of untethered galaxies is a result of the {\em acceleration} of the expansion of the universe, but not the {\em expansion} itself. Galaxies released from rest in accelerating cosmologies will recede outward, and galaxies released in decelerating (but expanding) cosmologies will move inward, often shooting past the observer and ``rejoining the Hubble flow''\footnote{For a detailed discussion of what it means to ``rejoin the Hubble flow'' see \citet{Barnes2006}.  Here we are referring in particular to their definitions 1 and 3.} in the opposite direction. \citet{Davis2003} argued such dynamics are consistent with the general relativistic description of expanding space; this expansion alone is not associated with any force on free particles, but accelerating expansion is. 

In the case of a relativistic velocity of recession, velocity corrections result in subtle but important modifications to the radial dynamics of these tethered galaxies. With the inclusion of these relativistic effects, a tethered galaxy released from rest in an expanding, {\em non-accelerating} cosmology will still experience an apparent  acceleration.  It will accelerate inward {\em against} the direction of the Hubble flow \citep{Whiting2004,GronElgaroy2007}. Although these relativistic dynamics are well accepted in the literature, the physical interpretation of these results has undergone significant debate. Papers such as \cite{Whiting2004} and \cite{Peacock2008} argue that the inward motion of tethered galaxies released from rest in expanding universes is qualitatively inconsistent with the picture provided through expanding space. However, \citet{Barnes2006}, \citet{GronElgaroy2007}, and \citet{Francis2007} use the radial motion of free particles from the relativistic geodesic equation to argue that expanding space provides a natural description of the motion of a free particle, as long as relativistic effects are carefully considered.

To shed light on the physical consequences of expanding space, we build upon the body of work provided through the tethered galaxy experiment to explore whether a physical tether could be used to hold a galaxy stationary at the Hubble radius (the distance at which $v_{\text{rec}} = c$).  We use the simplest case of a universe that neither accelerates, nor decelerates -- the empty universe. Through this paper, we interpret this thought experiment through the lens of several different reference frames, in order to separate events and effects from the coordinates that describe them. As we shall see, the definition of ``stationary'' itself is ambiguous, and what has zero velocity in one coordinate system is not necessarily stationary in another. 

Through Section II, we provide an overview of the global properties of the empty universe. The empty universe does not experience any acceleration or deceleration, and comoving particles recede with a constant velocity. We use this feature to reaffirm that expanding space, when disentangled from acceleration, is not associated with any force. In Section III we demonstrate the relativistically corrected tethered galaxy system of \citet{Davis2003} predicts an inward acceleration in the empty universe, and that this inward motion, as a relativistic velocity effect, needs no force, or spacetime curvature, to initiate it. In Section IV we extend this tethered galaxy system to connect a central observer to a galaxy which lies on the Hubble radius. In order to fix the position of a galaxy at the far end of our tether, we must supply an inward peculiar velocity equal to $c$. Our inability to hold a galaxy beyond the Hubble radius stationary is seemingly in contrast with the conclusion that expanding space is not associated with any force, developed in Section II. 

We resolve this apparent contradiction by expressing this ``Hubble tether'' system in the {\em inertial frame} of the central observer. By doing this, we show a tether that is stationary at a constant time in FLRW coordinates must be defined using points of varying inertial velocity. A tether along a constant FLRW time-slice that is longer than the Hubble radius is impossible to construct, as it requires $\vp > c$ even though there is no {\em Newtonian-like} force from expanding space. We express the inertial velocity profile of this tether, and show it restricts the length of objects to within the Hubble sphere, even in the limit of an empty universe. 

\section{The Canvas of Empty Space}

\subsection{General FLRW metric and expansion}

The expansion of the universe is most commonly described within the framework of the Friedmann-Lema\^{i}tre-Robertson-Walker (FLRW) metric, in which galaxies tracing the flow of expansion maintain a constant comoving coordinate ($\chi$). For a constant comoving coordinate, the amount these galaxies drift apart is proportional to scalefactor ($R$), such that their distance from the origin is $D=R\chi$ \citep{Weinberg2008}.

More fully, the standard FLRW metric is defined,
\begin{equation}
    ds ^2 = -c^2 dt^2 + R(t)^2 \left[d\chi^2 + S^2_k (\chi) ( d\theta^2 + \sin^2\theta d\phi^2)\right], 
\end{equation}
for interval $ds$, speed of light $c$, cosmological time $t$, comoving separation $d\chi$, and scalefactor $R$ (with dimensions of distance), expressed in spherical coordinates \citep{Peacock1998}.  We are only interested in radial distances, so $d\theta=d\phi=0$, leaving only $d\chi$. Thus we can see that the distance defined by $D=R\chi$ represents the distance between the origin and a comoving coordinate $\chi$ at a constant `cosmological time' ($t={\rm constant}$ or $dt=0$; where $t$ in FLRW also corresponds to wristwatch time for comoving observers).

Differentiating this distance with respect to cosmological time gives two components to the total velocity, $\vt$,
\begin{equation}
    v_{\rm t} = \frac{dR}{dt}\chi + R\frac{d\chi}{dt} = \vrr+\vp.
\end{equation}
Here the recession velocity, $\vrr$, comes entirely from the change in scalefactor $R$, and the peculiar velocity, $\vp$, comes entirely from a change in comoving coordinate.  Note that when you define the Hubble parameter to be $H\equiv \dot{R}/R$ (where the overdot represents differentiation with respect to FLRW time $t$), you recover the Hubble-Lema\^{i}tre law, $\vrr=HD$.  To achieve a total velocity of zero, we require the peculiar velocity to balance the recession velocity, $\vp = -\vrr$. 

None of the above requires general relativity, it is all determined by the FLRW metric, which is the general metric for a homogeneous, isotropic universe. However, in order to derive how scalefactor changes with time, and how objects move through comoving coordinates, we do need a theory of gravity. The Friedmann equation is the solution to general relativity with which we can calculate the evolution of the universe,  
\begin{equation}
\left(\dfrac{\dot{a}}{a}\right)^2 = H_0 ^2 \left[\Omega_{\rm K} a^{-2} + \Omega_{\rm M} a^{-3} + \Omega_{\rm R} a^{-4} + \Omega_\Lambda\right],
\end{equation}
where for convenience we have normalised the scalefactor to its value at the present day (subscript zero), $a\equiv R/R_0$, so $D=R_0a\chi$.  Each $\Omega$ represents a different component of the universe, with $\Omega_{\rm M},\; \Omega_{\rm R},\; {\rm and}\; \Omega_\Lambda$ representing respectively matter, radiation, and cosmological constant; and $\Omega_{\rm K} = 1-\Omega_{\rm M}-\Omega_{\rm R}-\Omega_\Lambda$ relating to the curvature of the universe. 

\subsection{Metrics of the Empty universe}

\subsubsection{Empty FLRW Metric}

In exploring what it means to say ``space is expanding,'' it is important to isolate the physical consequences of expanding space from those of accelerating (or decelerating) expansion. In order to disentangle these effects we make use of the cosmology where no such accelerations are experienced: the empty universe. If apparent accelerations arise in the empty universe, we know we must attribute them to whatever it means to say space is expanding, as there are no other accelerating features available. We therefore use this section to provide a brief overview of the global properties of the empty universe. 

Using the Friedmann equation with $\Omega_{\rm M}=\Omega_{\rm R}=\Omega_\Lambda=0$, we find $\dot{a}=H_0$; as expected, the expansion rate of the empty universe does not change with time.  So the scalefactor of this empty universe follows,
\begin{equation}
a(t) = H_0 t. \label{eq:at}
\end{equation}
For a galaxy at a comoving coordinate $\chi$, we may use Eq.~\ref{eq:at} to calculate its proper distance as a function of time, yielding,
\begin{equation}
    D(t) = H_0 t R_0 \chi.
\end{equation}
Comoving test particles in the expanding, empty universe recede with a constant velocity $H_0 R_0\chi$ throughout expansion. Such comoving particles are not accelerated by any force or tension from uniformly expanding space. This provides the backdrop of the empty universe in which we define the Hubble tether system.

\subsubsection{Milne Universe}

\begin{figure*}
    \centering
        \includegraphics[width=8.6cm]{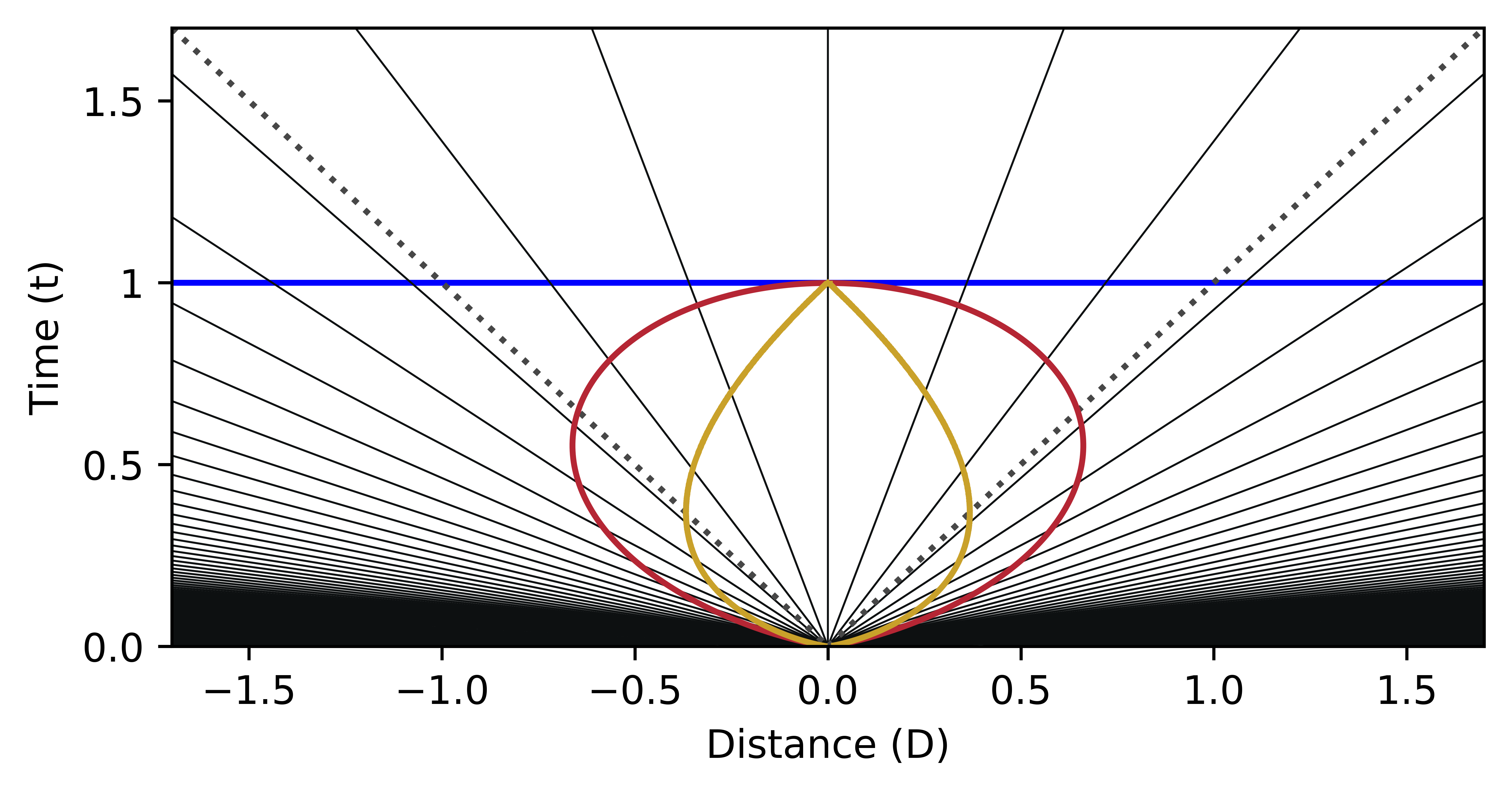}
        \includegraphics[width=8.6cm]{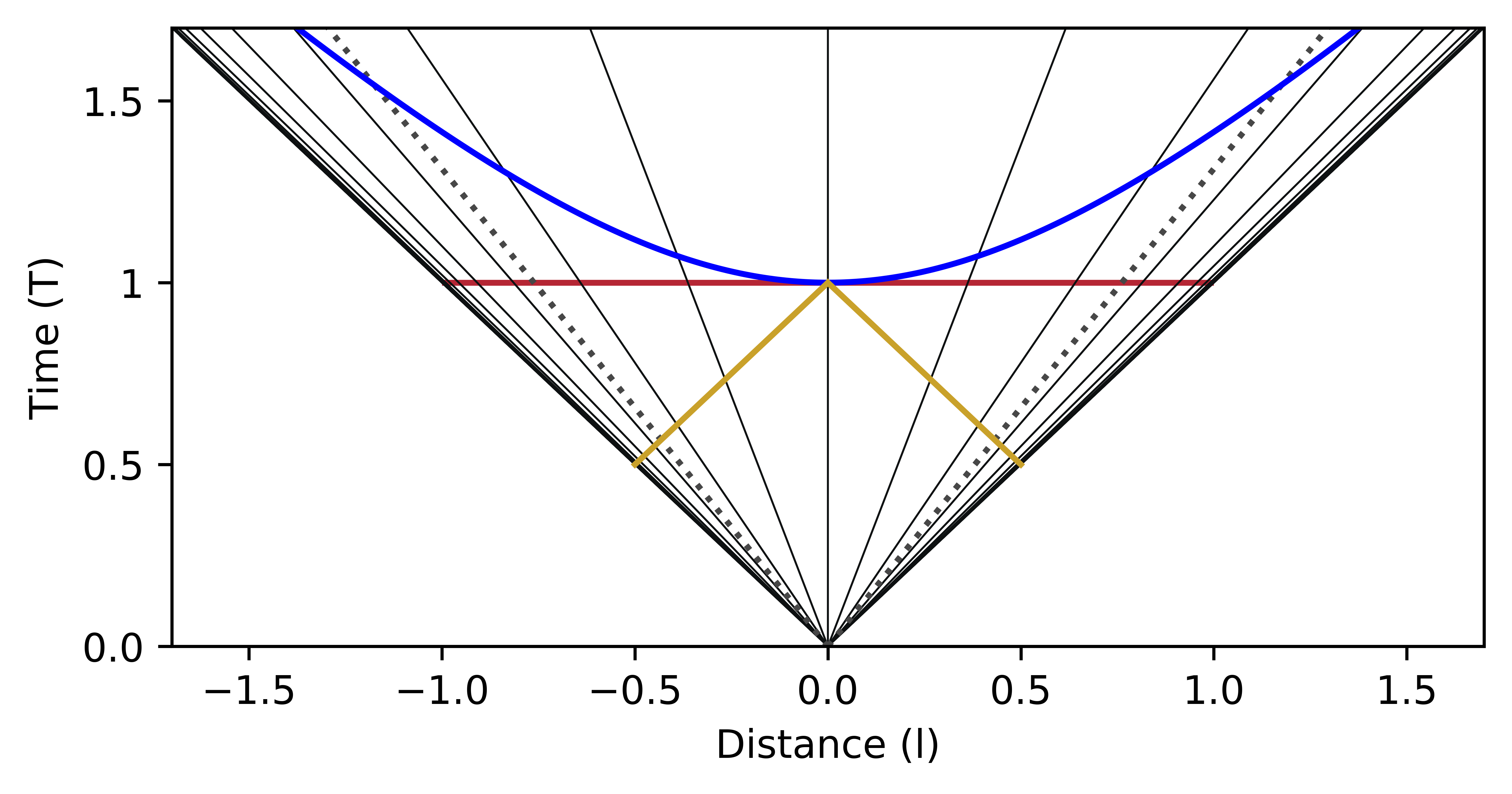}
        \caption[FLRW universe vs Milne universe]{Spacetime diagrams of the empty universe expressed in normalised FLRW coordinates (with $c = 1$) on the left, and using Minkowski coordinates as defined by a Milne observer on the right.  The same lines appear in both plots, the only difference being the coordinate system expressing them.  A constant FLRW time slice is shown in blue, while the inertial observer's constant time slice is in red.  Each point along the blue line corresponds to an observer that has seen the same time pass since the beginning of the universe as the observer at the origin.  Each point along the red line corresponds to a comoving particle with the same {\em inertial} time as our central observer at $D = 0, \ t = 1$. We populate this cosmology with equally spaced comoving particles in FLRW coordinates (black), which recede with a constant velocity. These particles occupy the entire area of this spacetime diagram in FLRW coordinates, or the future light cone in Minkowski space.  Note that the density of particles is homogeneous along a constant time slice in our FLRW representation, but increases with radius along a constant time slice in our Milne representation.  Test particles that lie on the Hubble radius recede with a constant FLRW velocity of $c$, denoted by the dotted line; in the Milne metric these same test particles have sub-luminal velocities of $v_{\rm M}= \tanh(1) \approx 0.762$. In both spacetime diagrams, we also include the past light cone of our observer at $D = 0, \ t = 1$ (denoted by the yellow line).}
        \label{fig:FLRWmilne}
\end{figure*}

An interesting alternative, but completely equivalent, description of the empty universe is provided through the Milne model. The Milne universe recasts the empty FLRW universe into the coordinates of a single, inertial observer, defined to be the origin of a Minkowski spacetime. By expressing the dynamics of particles from the inertial reference frame of our Milne observer (a globally inertial reference frame), we aim to resolve confusion resulting from how distances are defined in the spatially curved frame of the empty FLRW universe.\footnote{Note that even though the Milne universe can not be directly compared to the general FLRW cosmologies (with non-zero mass and energy), the empty universe it corresponds to is a convenient cosmology to test our assumptions of expanding space when disentangled from acceleration. The physical interpretation we assign to the features of expanding space in FLRW co-ordinates must be able to explain the observations of a Milne observer. If our assumptions do not hold in this simple case (with no accelerating features), then we can not expect them to hold in the more general case of cosmologies with accelerating/decelerating expansion.} The standard Minkowski metric is defined,
\begin{equation}
     ds^2 = -c^2 dT^2 + dl^2 + l^2 d \Phi^2,
\end{equation}
for radial distance $l$, inertial time $T$, and 3D angular coordinates $\Phi$. In the Milne universe, massless test particles spread from the big bang ($l = 0, \ T = 0$) with a constant velocity ($v_{\rm M} = l / T$) through this pre-existing Minkowski space. Any test particle can be chosen as our ``privileged'' inertial observer, and any choice will yield the same distribution of test particles, meaning the Milne universe has no unique centre \citep{Rindler1977}. While the Milne and empty FLRW universe describe the same cosmology, the key difference between these models lies in their choice of spacetime coordinates. The empty FLRW universe is explicitly defined such that the density of particles along a constant time-slice is homogeneous, which is satisfied when every particle agrees on their local, wristwatch time. Since the Milne and empty FLRW universe describe the same cosmology, the Milne universe shares this pattern of test particles. As seen in Figure~\ref{fig:FLRWmilne} however, this distribution of test particles is no longer homogeneous when measured along a constant Milne time-slice. Time in the Milne universe is defined using the inertial time of the privileged (but arbitrarily chosen) observer at the origin. Objects that are further away (and as such, have been receding with a greater velocity) undergo special relativistic length contraction from the reference frame of this inertial observer. This length contraction results in a density profile which increases with distance along our inertial Milne time-slice. The pattern of test particles in both models is exactly the same, we have only changed the coordinates used to represent them. We may directly transform the coordinates of the empty FLRW universe into the coordinates of our inertial Milne observer through the relation,
\begin{equation}
    l = ct \sinh\left(\chi\right); \quad
    T = t \cosh\left(\chi\right). \label{eq:FLRWMilneConversion}
\end{equation}
The empty FLRW and Milne universe are interchangeable models that describe the same pattern of test particles within the same, empty cosmology. However, as we shall develop in the next section, their differing choices of time-slice yield different interpretations of recession and expansion that we wish to explore. 

\subsection{The distribution of test particles in empty-FLRW vs Milne universes}
Physical processes are coordinate independent, but the language we use to describe these processes is often rooted in the coordinate system we use.  In order to highlight how using an inertial reference frame affects the description of the empty universe, consider a series of comoving test particles defined with an equal comoving separation in the FLRW metric. Along a constant FLRW time-slice (the slice where all test particles agree on the age of the universe), one would record a homogeneous test particle density extending over all space (Figure~\ref{fig:FLRWmilne}, left).

By contrast, when this is converted to the Milne universe, we find a density that increases with radius. As developed earlier, when we cast our empty universe in the reference frame of a privileged, inertial observer, special relativistic length contraction results in an increasing test particle density with distance (Figure~\ref{fig:FLRWmilne}, right). This effect causes all test particles (and by extension, the infinite FLRW universe) to be contained within the finite distance of the future light cone of the big bang in the Milne universe.

Our choice of time-slice has particularly significant consequences when considering the recession of a galaxy on the Hubble radius. In the FLRW metric objects can recede with velocities exceeding $c$ without any relativistic violation. This superluminal recession occurs outside the inertial reference frame of any observer, resulting in no special relativistic contradiction \citep{Murdoch1977,Davis2004}.  As such, the more distant the galaxy, the faster it can recede without limit. However, the Milne universe (cast on a flat, Minkowski spacetime) is subject to special relativistic constraints, meaning recession greater than $c$ is not possible in this coordinate system.

The fact these two coordinate systems demonstrate such differing interpretations of recession for the same universe highlights the fact that velocity is a coordinate dependent property. A coordinate-independent feature would be the four-velocity, but that is not the velocity that appears in Hubble's law, and more broadly it is not the standard velocity invoked to interpret cosmological recession. Importantly, dimensionless observables have to remain true regardless of coordinates, such as redshift.

By contrasting the description of cosmological recession in the empty universe expressed in the framework of the Milne and FLRW coordinate systems, we aim to isolate the properties of expansion that are fundamental physical characteristics, from those that are a consequence of our coordinate representation. 

\section{The Tethered Galaxy System}

With the backdrop of the empty universe now firmly established, we use this section to explore the tethered galaxy system as a framework for understanding the features of expanding space. In this section, we closely follow the tethered galaxy system outlined in \citet{Davis2003}, replicating their results in the limit of non-relativistic velocities. We then apply a relativistic correction provided (but not directly used) in their paper.

In generating these results, we make use of the fact that (in the limit of non-relativistic velocities) momentum with respect to the local comoving frame decays as $1/a$ \citep{MTW1973}. We demonstrate that such scalefactor arguments (when relativistic effects are included) display the same behaviour as the full general relativistic derivation, which you can find in \citet{GronElgaroy2007}.

\subsection{Non-Relativistic Limit}

\begin{figure}[h!]
    \centering
        \includegraphics[width=8.6cm]{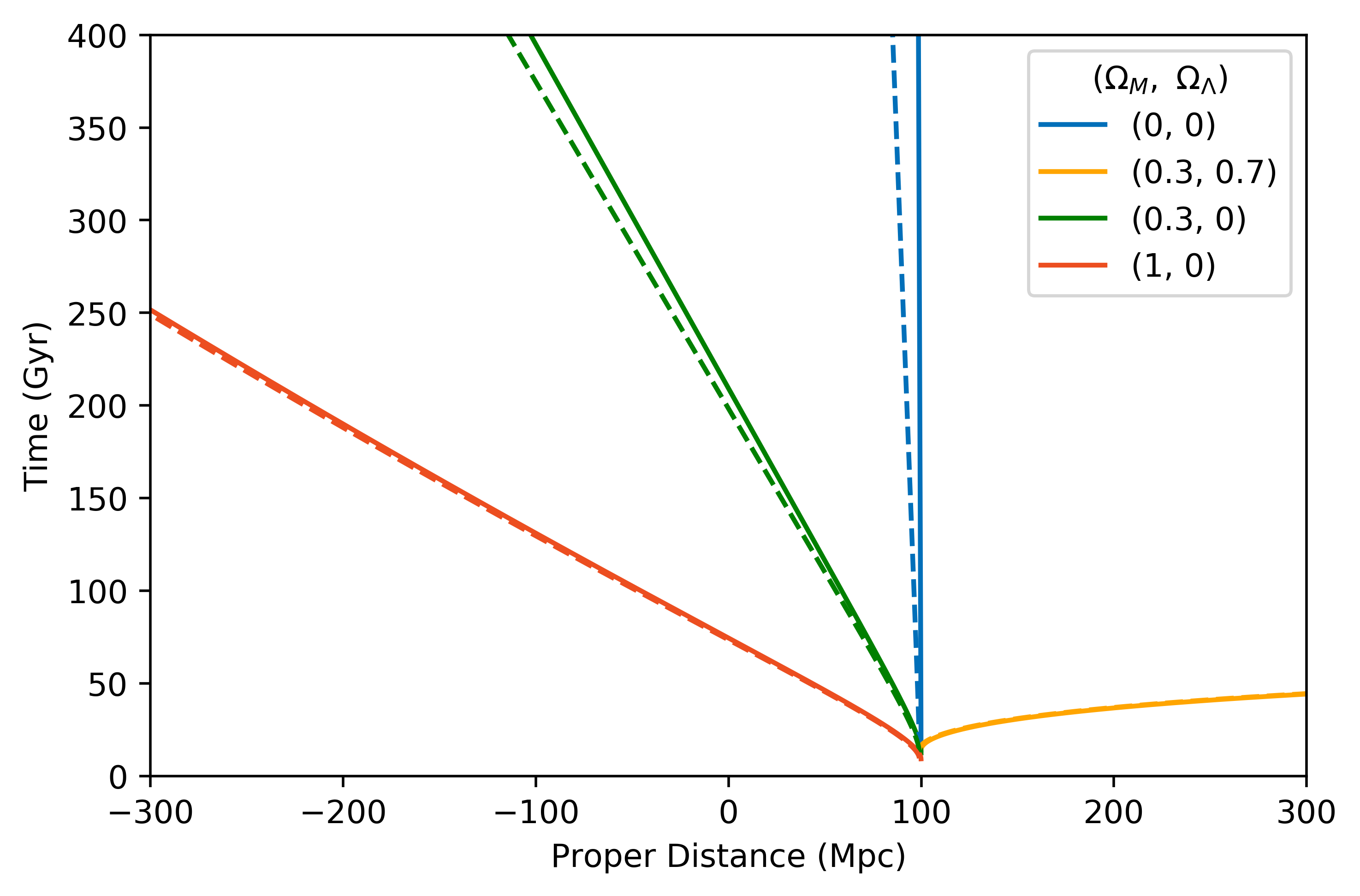}
        \caption[Dynamics of tethered galaxies released in expanding cosmologies (non-relativistic)]{
        The dynamics of a tethered galaxy in four different cosmologies as specified in the legend.  The non-relativistic solution is given as solid lines \citep[replicating Fig.~2 of][]{Davis2003}, while the relativistic solutions are given as dashed lines. In the non-relativistic case the tethered galaxy moves inward for cosmologies with decelerating expansion, outward for cosmologies with accelerating expansion, and stays at a constant distance for cosmologies with constant expansion.  In contrast, the relativistic solution shows a small inward co-ordinate acceleration in the universe with constant expansion. We clarify that the trajectories of these untethered galaxies correspond to geodesic motion. 
        This is a spurious acceleration arising because of our coordinate choice, and is not supplied by any force. 
        This acceleration is a co-ordinate acceleration (i.e, $\frac{d^2D}{dt^2} \neq 0$), and does not correspond to any relativistic 4-acceleration.
        This release distance ($100$ Mpc) is similar in magnitude to the BAO scale ($\sim 150$ Mpc), however since the BAO feature corresponds to a comoving standard ruler, rather than a fixed physical size, one would not expect peculiar velocity decays to have an effect.}  
        \label{fig:Davis}
\end{figure}

In an expanding universe, distant galaxies undergo cosmological recession, $\vrr = HD$. In the tethered galaxy system, we imagine that, as an initial condition, an inward peculiar velocity is supplied, which is of an equal magnitude to this recessional velocity $\vpz=-\vrz$. As such, the proper distance $D$ of the galaxy from the observer is initially fixed and the initial total velocity $\vtz=0$ (as if it had been physically tethered out of the Hubble flow). Equivalently,
\begin{equation}
    R_0 \dot{\chi}_0 = -\dot{R}_0 \chi_0. \label{eq:R0chi0}
\end{equation}
The peculiar momentum of this tethered galaxy ($p = m \vp$) will decay with respect to the comoving background frame as $1/a$ \citep{MTW1973}. In the non-relativistic limit, the peculiar velocity tethering this galaxy decays by this same factor,
\begin{equation}
   \vp = \frac{\vpz}{a}. \label{eq:vpa}
\end{equation}
Combining this result with Equation~\ref{eq:R0chi0} gives,
\begin{equation}
    R \dot{\chi} = \frac{-\dot{R}_0 \chi_0}{a}.
\end{equation}
Recalling $R=R_0a$ and integrating from $\chi_0$ to $\chi$ and $t_0$ to $t$ gives,
\begin{eqnarray}
    D &=& R \chi_0 \left(1- H_0 \int_{t_0}^t \frac{dt}{a^2} \right),  \label{eq:DTimeDeriv} \\
    &=&  R \chi_0 \left(1- H_0 \int_{1}^a \frac{da}{\dot{a}a^2} \right). \label{eq:D}
\end{eqnarray}
One can compute this integral numerically for any arbitrary cosmology by using Friedmann's equation to supply $\dot{a}$.  We calculate this for the same set of cosmologies described in \citet{Davis2003}, directly replicating their results as the solid lines in Figure.~\ref{fig:Davis}.  The empty universe is shown by the blue line, and you can see that in the non-relativistic limit a galaxy that starts at rest ($\vt=0$) in the empty universe remains at rest.

\subsection{Relativistic Limit}

As the peculiar velocity of tethering enters relativistic regimes, velocity corrections result in subtle but important modifications to the dynamics of untethered galaxies \citep{Whiting2004,GronElgaroy2007}. For a relativistic peculiar velocity, we must use the relativistically adjusted momentum formula,
\begin{equation}
    p = \gamma m \vp,\label{eq:mom}
\end{equation}
where $\gamma=(1-\vp^2/c^2)^{-1/2}$.  Since $p \propto 1/a$, we obtain,
\begin{equation}
\vp = \frac{\gamma_0\vpz}{\sqrt{a^2 + \gamma_0^2 \vpz^2 / c^2}}. \label{eq:relvp}
\end{equation}

\begin{figure}
    \centering
        \includegraphics[width=8.6cm]{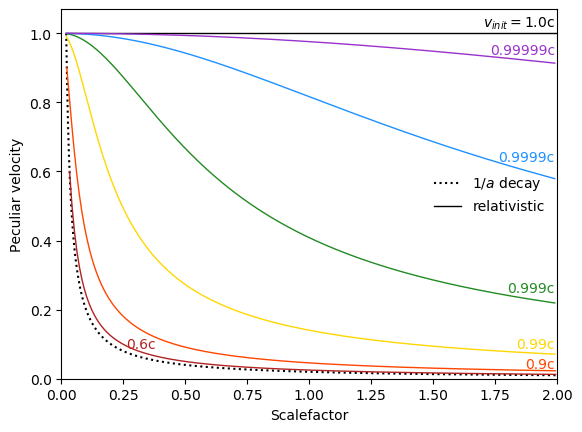}
        \caption[Velocity decay]{How peculiar velocities decay as the universe expands, showing the continuum between relativistic and non-relativistic behaviour.  The non-relativistic solution has $\vp=v_{\rm p,0}/a$ and is shown as a dotted line.  The relativistic solution is given by Eq.~\ref{eq:relvp} and is shown as solid lines for 7 different initial velocities as labelled. The initial scalefactor is $a=0.02$, and the $v_{\rm init}=0.9c$ and $0.6c$ cases begin from the same scalefactor as the dotted line for easy comparison. } 
        \label{fig:velocitydecay}
\end{figure}

Figure~\ref{fig:velocitydecay} shows this results in a continuum of behaviours between photons (whose peculiar velocities do not decay) to non-relativistic massive particles (whose peculiar velocities decay as $1/a$).  In between these extremes, relativistic massive particles have velocities that decay more slowly than $1/a$.  Interestingly, while the wavelengths ($\lambda$) of photons are redshifted by the expanding universe, so are the de Broglie wavelengths of massive particles. In both cases $\lambda \propto a$, however for massive particles the redshifting of their de Broglie wavelength manifests itself as a decrease in peculiar velocity, while for photons it just results in a longer wavelength.

Using this relativistically adjusted peculiar velocity function to replace Equation~\ref{eq:vpa} in the derivation of Equation~\ref{eq:D} yields,
\begin{equation}
    D=R \chi_0 \left(1- \gamma_0 H_0 \int_{1}^a \frac{da}{\dot{a}a\sqrt{a^2 + \gamma_0^2 \vpz^2 / c^2}} \right). 
    \label{eq:Drel}
\end{equation}
When applied to the same cosmologies as in the previous section this generates the dashed lines showing the relativistic trajectories of untethered galaxies in Fig.~\ref{fig:Davis}.

It is clear that in the full relativistic solution, we find that tethered galaxies released from rest in the expanding, empty universe experience an apparent inward acceleration. Interestingly, the tether does not get stretched apart by expanding space as we might initially assume, but rather collapses. This inward acceleration has been described by others in the literature, notably \citet{Whiting2004} and \citet{GronElgaroy2007}. 

\subsection{Why peculiar velocities decay}
There is a simple physical explanation of why peculiar velocities decay.  Imagine a vertical worldline in the FLRW spacetime diagram of Figure~\ref{fig:FLRWmilne}. (Such a worldline is only physically possible within the Hubble sphere, although that detail is not important for this argument.)  A vertical worldline means the object is stationary in these coordinates ($\vt=0$), and therefore it must have an inward peculiar velocity equal and opposite to the outward recession velocity ($\vp=-\vrr$).  As time goes on, the object gets overtaken by comoving galaxies that are moving more and more slowly (the solid black lines become more vertical).  So to stay at a constant distance, the object's peculiar velocity has to decrease.  This example shows that peculiar velocity decay is just a bookkeeping exercise, arising because we have split our velocity into two components.  The total velocity $\vt$ is zero and stays zero --- no accelerations have been applied. 

The surprising thing is that in the relativistic case, the peculiar velocity does not decay fast enough to keep the object at $\vt=0$.  When we apply the relativistic correction, the inward peculiar velocity decays at a reduced rate, generating an inferred inward acceleration as in Figure~\ref{fig:Davis}. This inward acceleration of tethered galaxies released in the empty universe is just a result of relativistic addition of velocities, and does not require any Newtonian-like force. 

\begin{figure*}
    \centering
        \includegraphics[width=8.6cm]{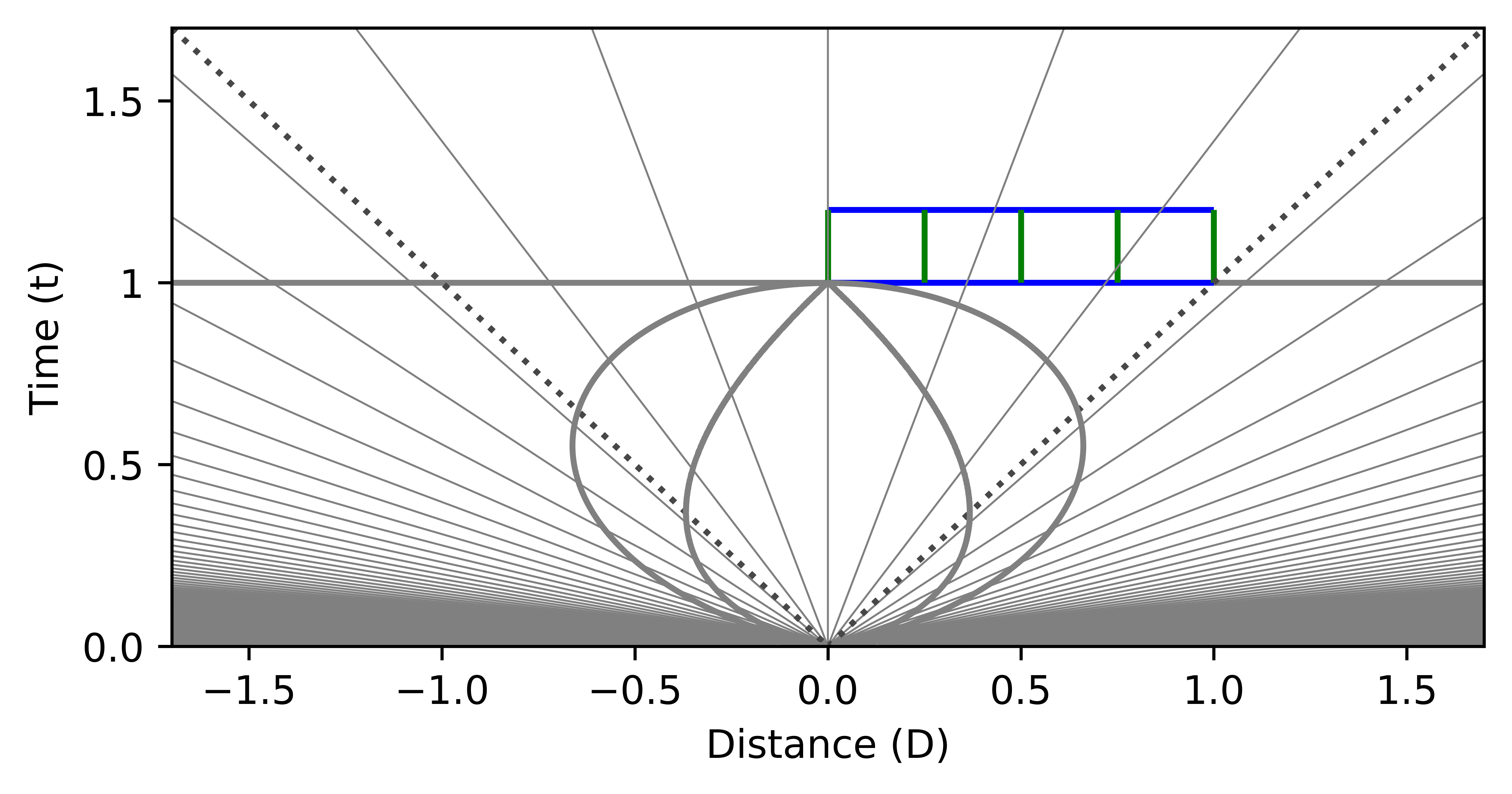}
        \includegraphics[width=8.6cm]{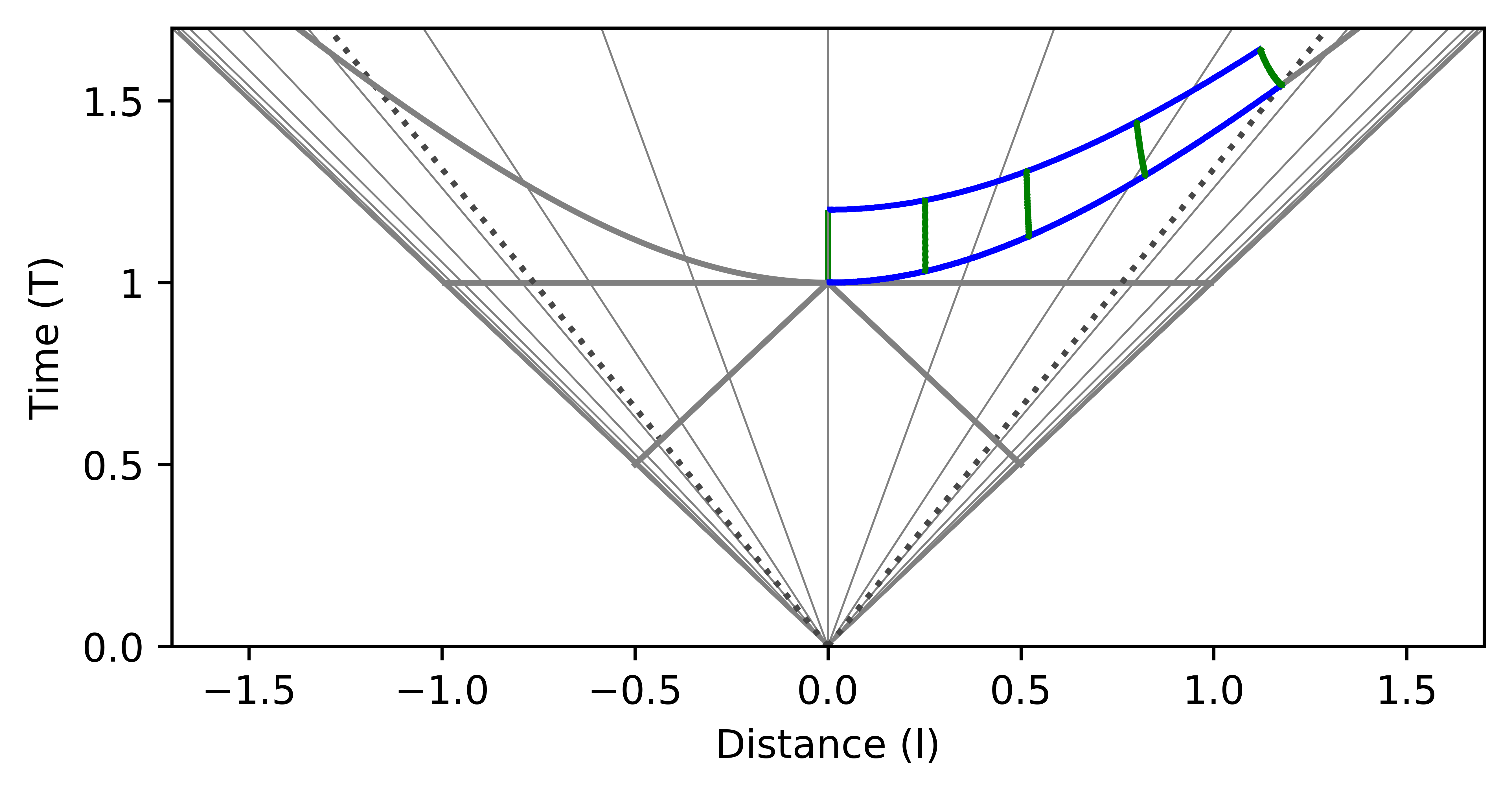}
        \caption[Time evolved Hubble tether]{Evolution of Hubble tether system whose reference frame is defined to coincide with a constant FLRW timeslice from $t=1$ to $t = 1.2$. As a massless test galaxy is tethered out of the Hubble radius, the FLRW distance of the galaxy from the central observer is fixed. This causes the tethered galaxy to follow the green, vertical worldline centred around $D = 1$. Similarly, all of the elements of our tether have fixed FLRW positions, and follow parallel vertical worldlines. After following these vertical worldlines, these points define a tether of the same length at the later time $t = 1.2$ (given by the later blue line).  In the Minkowski coordinates of the Milne universe the Hubble tether follows a hyperbola.  In order to stay the same length in FLRW space, it must shrink in Minkowski space.  The green lines are sloped inwards, indicating the velocities through this inertial frame for the different parts of the tether.  When the end is on the Hubble sphere, the inward velocity needed for that end to stay at a constant FLRW distance is $c$. This is why you cannot tether a particle beyond the Hubble sphere.}
        \label{fig:FLRWHubbleTether}
\end{figure*}

\section{The ``Hubble Tether'' System}
\subsection{Constructing a ``Hubble Tether''}
Now we take the relativistic limit to the extreme. In order to evaluate the physical consequences of expanding space in maximally relativistic regimes, we consider a massless test galaxy that lies on the Hubble radius of some central observer, which is held in place by a tether spanning this distance. We imagine the job of designing this ``Hubble tether'' system has been delegated to a student of cosmology, who defines the tether to lie along a constant FLRW time-slice (given by the horizontal blue line of the left panel of Figure~\ref{fig:FLRWmilne}). To achieve this, they require every point along this tether be defined with massless test particles, that are at rest in FLRW coordinates (i.e. $v_{\text{t}} = 0$), and stay at rest.\footnote{The term ``at rest'' relative to the Hubble expansion is often used to refer to comoving galaxies.  Here we use it to mean at rest with respect to the FLRW coordinates.} This means each point has a peculiar velocity equal but opposite to the recessional velocity at that point, and they are adjusted as necessary to stay that way, yielding the spacetime diagram of Figure \ref{fig:FLRWHubbleTether} (left).\footnote{For a more observer-centred approach to constructing this system, one could imagine each element along our tether uses radar ranging to fix their local positions \citep[following the treatment of][]{Geraint2008}. In the limit of small distances the empty FLRW universe is spatially flat, meaning each element could accelerate/decelerate as necessary to remain stationary with respect to their adjacent elements, defining our stationary tether.} Note that this is different to defining a ``rigid'' rod. We choose to analyse this system without referring to rigidity, due to the technical challenges associated with defining such a property in a relativistic context \citep[Ch.~16 and Ch.~2 of][respectively]{panofsky1964,landau1971}. In particular, rigidity may seem to imply there are forces internal to the rod holding it in place, raising the issue of faster-than-light communication to transmit such forces. Our ``rod'' or ``tether'' is simply defined to be a series of adjacent particles, of fixed separation in the frame defined by FLRW co-ordinates. Whether or not these particles move away from or toward each other (in a Minkowski frame) when allowed to move freely will determine whether a rod or tether would need internal forces in order to hold itself together or stop itself collapsing. 

What does such a system look like from the reference frame of an inertial observer? When we cast this system in the inertial co-ordinates of our Milne observer (an observer at rest along $D = l = 0$), the Hubble tether system (defined over a constant FLRW time-slice) now stretches across a hyperbolic slice. Using the Milne-FLRW conversion of Equation~\ref{eq:FLRWMilneConversion}, we can express the time evolved Hubble tether system in the inertial coordinates of the central observer, yielding Figure \ref{fig:FLRWHubbleTether} (right).

By casting our time evolved Hubble tether system in the coordinates of an inertial observer, we find the fixed elements that define a constant tether in FLRW coordinates must be constructed using a range of inertial velocities. While the observer end of our tether follows a stationary vertical worldline in the Milne universe (about $l = 0$), the points extending along this tether require accelerating kinematic profiles to maintain their FLRW position. It is this requirement that the tether be at rest against the frame defined by the standard FLRW coordinate system, which does not correspond to the inertial reference frame of any observer, that causes this tether to collapse by construction.

\begin{figure*}
    \centering
        \includegraphics[width=8.6cm]{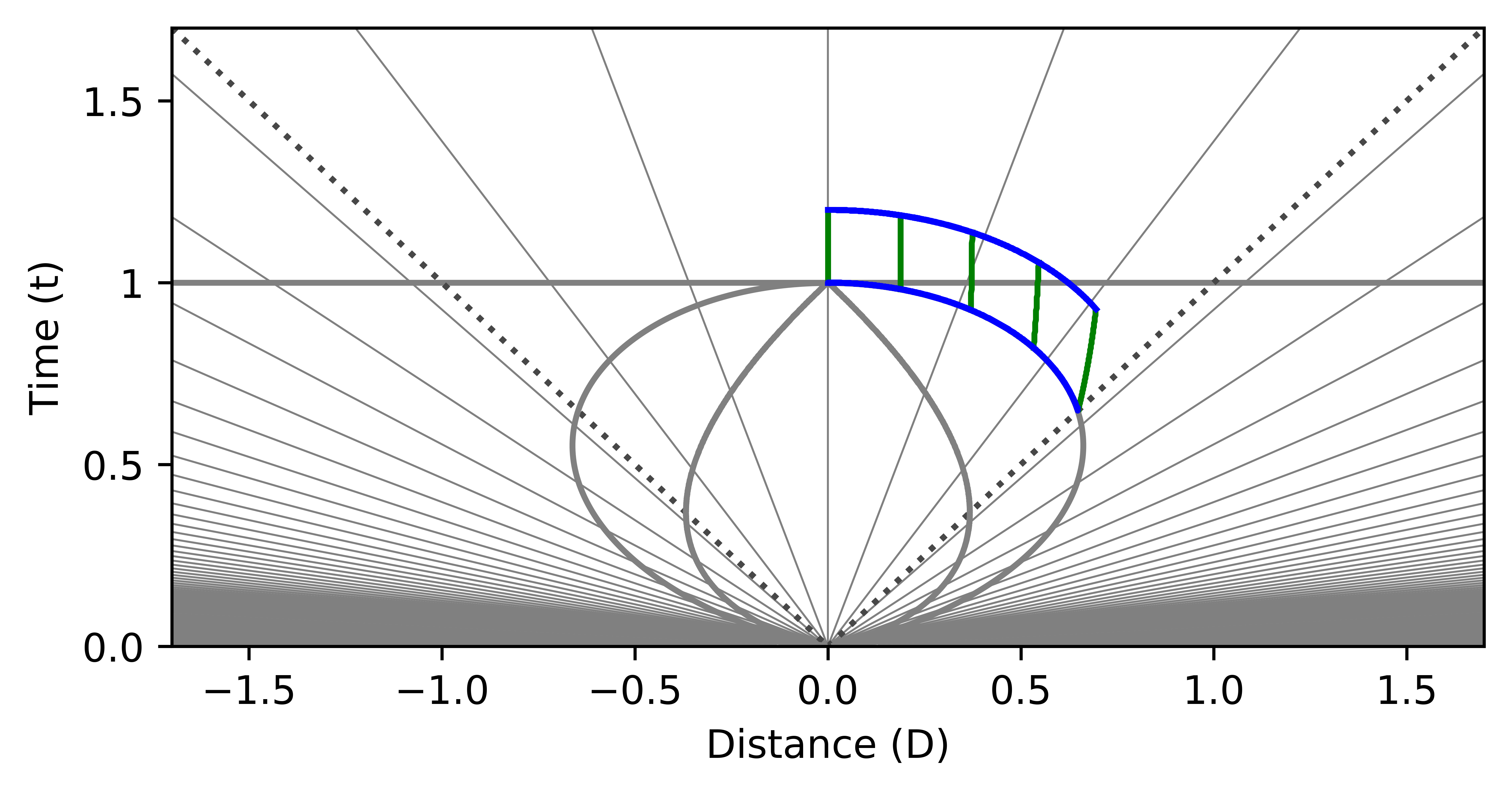}
        \includegraphics[width=8.6cm]{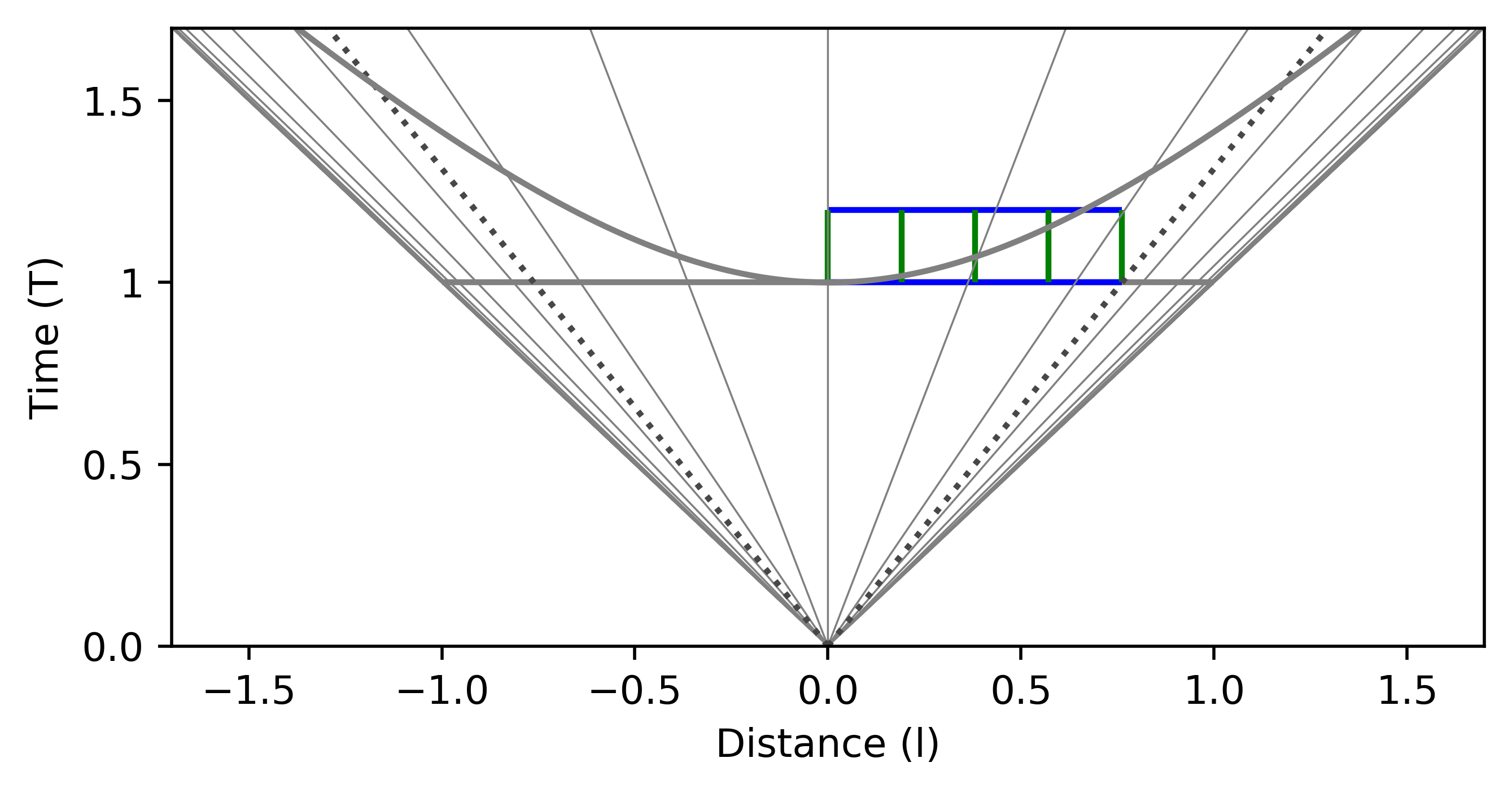}
        \caption[Time evolved Milne Tether]{Alternative view of a ``stationary'' tether, this time defined to be stationary in the coordinates of the Milne model. The inertial tether, when expressed in FLRW coordinates, actually has its distant end embedded in the past--- in a universe with a higher test particle density. Additionally, in order to remain at a fixed distance as measured by our inertial observer, the elements defining this tether move along curved worldlines in FLRW co-ordinates. As a result, our student assessing the tether design would infer the particles defining this tether are moving apart.}
        \label{FLRWHubbleTetherConstMilne}
\end{figure*}

In contrast, if we were to define the tether at a constant length in the arguably more natural reference frame of an inertial observer, we arrive at the spacetime diagram of Figure \ref{FLRWHubbleTetherConstMilne}. Note that when the elements of our tether are defined with constant separations in this inertial reference frame, our object does fulfil the condition of Born rigidity. When we cast our inertial tether back into FLRW coordinates, we find the far end is embedded in the past (which has a higher test particle density). Additionally, in order to remain at a constant position with respect to our inertial observer, the elements defining this tether must move outward in FLRW coordinates. This shows that a tether which is at rest in the inertial Minkowski frame, would appear to lengthen in the FLRW frame, even though no stretching is occuring! Conversely, this shows that a tether which is ``at rest'' in FLRW coordinates in not truly at rest in the inertial frame of any observer.

\subsection{Inertial Velocities along tether}
While the previous section allows us to describe the behaviour of our constant tether in the empty universe, a key part of our question still remains--- What, if anything, is significant about a \textit{Hubble} tether? Is there anything significant about the Hubble sphere in the description of our tethered galaxy system? To answer this we return to our constant FLRW tether, and explore the inertial velocities of the elements along this length (with an FLRW velocity of 0) from the central observer to the Hubble sphere. To develop this profile, we constructed two lines of constant length in FLRW coordinates, separated by some infinitesimal time $dt$. This corresponds to the time evolved Hubble tether system of Figure \ref{fig:FLRWHubbleTether}, in the limit of an infinitesimal time separation. Using the FLRW-Milne relation of Equation~\ref{eq:FLRWMilneConversion}, we solved for the initial ($l_{\textrm{init}}, T_{\textrm{init}}$) and time separated ($l_{\textrm{final}}, T_{\textrm{final}}$) Milne coordinates of the elements along these horizontal FLRW lines. This allows for a numerical evaluation of the Milne velocities required to yield $v = 0$ in FLRW coordinates with the simple relation:

\begin{equation}
    \frac{dl}{dT} = \frac{(l_{\textrm{final}} - l_{\textrm{init}})}{(T_{\textrm{final}} - T_{\textrm{init}})}.
\end{equation}

\begin{figure}[h!]
    \centering
        \includegraphics[width=8.6cm]{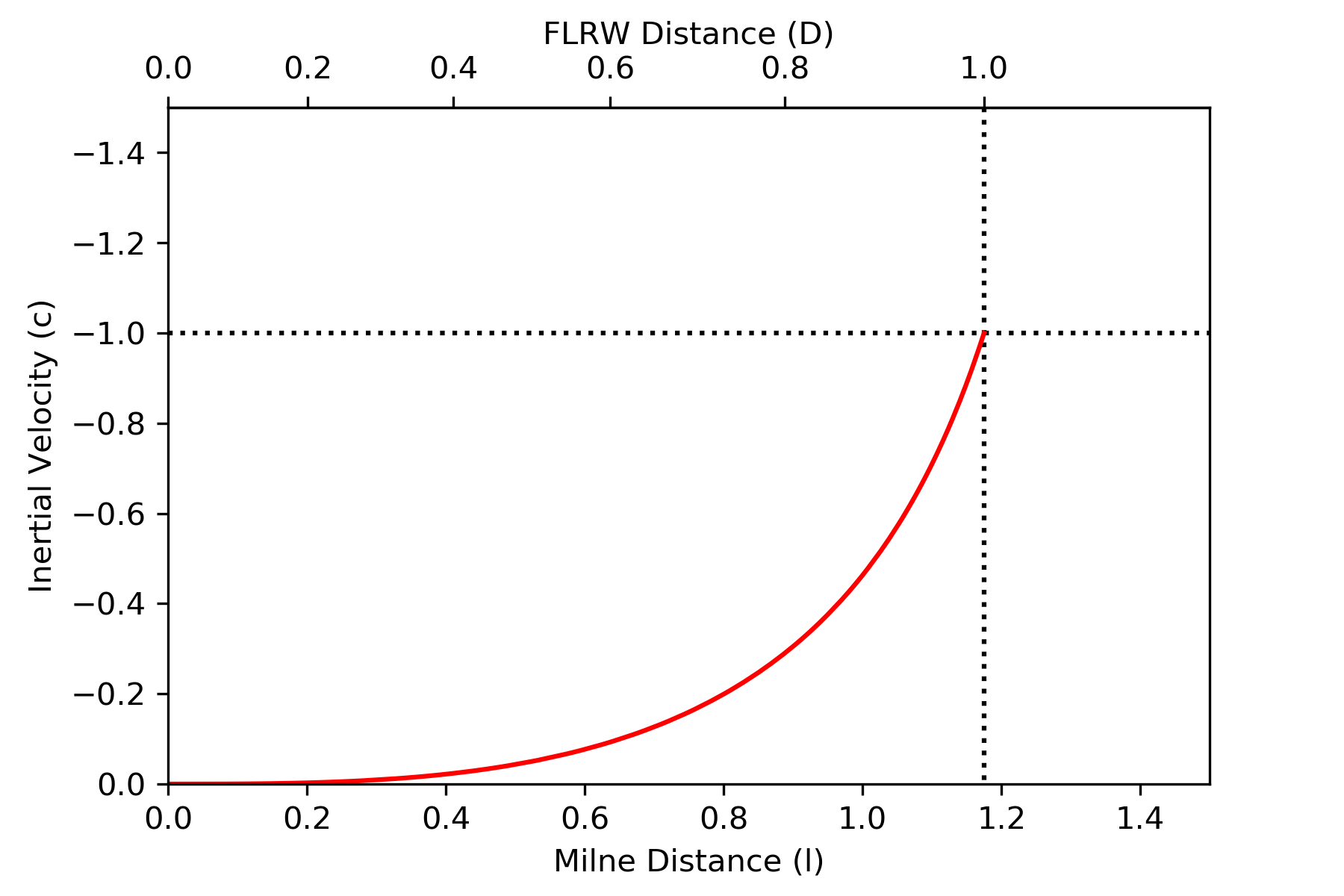}
        \caption[Profile of peculiar velocities required FLRW tethering]{Profile of inertial velocities required to construct Hubble tether system along $t = 1$, as a function of both Milne and FLRW distance. As one follows a hyperbolic time slice of constant FLRW $t$, the peculiar velocities required to tether these points with a fixed FLRW distance increases dramatically. At the end of our Hubble tether (along the dotted line of $l \approx 1.175$), an inertial velocity of precisely $c$ is required to tether this system in FLRW coordinates.}
        \label{fig:InertialVelocity}
\end{figure}

The inertial velocity profile of particles along the length of our Hubble tether is given as in Figure~\ref{fig:InertialVelocity}.

We can see the Hubble sphere represents a fundamental limit to how far we can define such a chain of particles, regardless of any rigidity constraints. In order for a tether that extends to the Hubble radius to maintain a constant length in FLRW coordinates, we must supply an inward inertial velocity equal to $c$, beyond which is relativistically forbidden. Imagine we define a series of observers, each of which have rockets to change their inertial velocity. As we attempt to extend this chain of observers along a constant FLRW time-slice,  the inertial velocities required to fix their FLRW position increases dramatically. As we extend this chain to a point on the Hubble radius of the central observer, we must supply an inertial velocity of $c$ to fix our FLRW position. The worldline of a galaxy on the Hubble radius is subluminal in Milne coordinates (as seen in Figure 1), but fixing the position of this particle in FLRW coordinates beyond the Hubble sphere requires superluminal motion through this inertial frame, which is why such a tether cannot be constructed.

\section{Discussion and Conclusion}

With the discovery of the accelerating expansion of the universe, the need for clarity in how we interpret this expansion has become increasingly important. The description of the expanding universe, and the interpretation provided through expanding space, fundamentally frame our approach to open problems in cosmology. By exploring the dynamics of a particle tethered out of the Hubble flow in the empty universe, we show such a particle experiences an inward acceleration. While such a result has been presented in the literature previously, our treatment highlights how this is well described as a relativistic velocity decay. Such dynamics do not require any force or acceleration from expanding space. 

The properties of expanding space appear to come into question again when considering the tethering of a galaxy on the Hubble radius (with a recessional velocity of $c$) in the empty universe. We find such a tether cannot be constructed, seemingly in contrast to the conclusion that expanding space does not stretch the matter it contains. This apparent contradiction is resolved when considering the effect of our choice of time-slice. By choosing the standard FLRW time-slice, we choose a global frame that does not correspond to the frame of any inertial observer. Simple relativistic transforms introduce accelerations where no such force exists, similar to the fictitious forces which arise in rotating reference frames. By enforcing the requirement that each element of our tether is at rest in the frame defined by the standard FLRW coordinate system (which does not correspond to any inertial reference frame), we require they accelerate in the reference frame of an inertial observer. It is this requirement that causes our tether to collapse, and not any force from the expansion of space. 

It is interesting that we cannot construct a Hubble tether which is stationary with respect to the global frame defined by the FLRW coordinate system because objects at $D>D_H$ cannot be held at a fixed FLRW distance -- they must move away, seemingly `tearing' any attempt to make a tether in this regime. Whereas within the Hubble sphere once you set up the initial condition of $v_t=0$ the tether does not rip apart, it collapses.  That behaviour is true for an empty universe, but in general it differs depending on what type of universe you live in.  Assuming a perfect fluid with equation of state $w=P/\rho$, \cite{GronElgaroy2007} determined the initial displacement necessary to observe inward acceleration, as a function of $w$,
\begin{equation} \chi_0^2 > - \dfrac{1}{2} (1 + 3 \omega) \left(\dfrac{c}{H_0}\right)^2. \label{eq:gronlim} \end{equation}
If $w=-1$ this corresponds to exactly the Hubble sphere, so you do not get inward acceleration.  However, in the empty universe we have been considering, for which $w=-1/3$ you get inward acceleration from any $\chi_0>0$.   Decelerating universes ($w>-1/3$) also obviously have inward acceleration regardless of the distance.

To return to the question we posed in the abstract, does this mean we cannot construct an infinitely long tether? If we define this tether along a constant inertial time-slice, there is no reason it could not extend to the edge of the Milne universe (the future light cone of the big bang). In such a case, the tether would follow the red inertial time-slice of Figure~\ref{fig:FLRWmilne}, right. As this constant tether extended further and further, the far end would extend further and further back in cosmological time ($t$). In order for this ruler to extend to the edge of the Milne universe (where particles have an inertial recessional velocity of $c$), the far end would be embedded at $t = 0$ in the FLRW universe.

While questions regarding the fundamental nature of expanding space may seem esoteric, it is interesting to consider the potential physical implications of relativistic peculiar velocity decays within our own universe. Super-clusters form filaments that can span up to 10 Gly, occupying up to 5\% of the observable universe \citep{Horvath2013}. Given the existence of structure at such large scales, it is interesting to consider whether analogues of proper tethers exist on cosmological scales. If systems with fixed proper distances do indeed exist on such scales, one would expect the relativistic decay of peculiar velocities would provide an increased inward ``kick'' to support a slightly higher growth rate for such structures than calculated in a purely Newtonian simulation. 

\begin{acknowledgements}
We thank Colin MacLaurin, Geraint Lewis, and Angela Ng for useful discussions. TMD is the recipient of an Australian Research Council Australian Laureate Fellowship (project number FL180100168) funded by the Australian Government. AG is the recipient of an Australian Government Research Training Program (RTP) Scholarship.
\end{acknowledgements}

\begin{appendix}

\section{Conversion from Milne to empty FLRW universe}

Here we demonstrate the Milne and FLRW universe are identical. Given a constant velocity of recession ($v_{\rm M} = \frac{l}{T}$) in conjunction with the conversion of Equation \ref{eq:FLRWMilneConversion}, 
we may express our special relativistic $\gamma$ in terms of our comoving space and time ($\chi, t$) coordinates,

\begin{equation}
    \gamma = \dfrac{1}{\sqrt{1 - v_m^2/c^2}},
\end{equation}

\begin{equation}
    \gamma = \dfrac{1}{\sqrt{1 - \left(\frac{ct \sinh(\chi)}{t \cosh(\chi)}\right)^2/c^2}},
\end{equation}

\begin{equation}
    \gamma = \cosh{\chi}.\label{eq:gammacosh}
\end{equation}
Substituting the time relation of Equation~\ref{eq:FLRWMilneConversion} and Equation~\ref{eq:gammacosh} 
into the Lorentz boosted Minkowski time coordinate ($T = \gamma T'$) yields,

\begin{equation}
    T' = \dfrac{t \cosh(\chi)}{\cosh{(\chi)}},
\end{equation}

\begin{equation}
    T' = t. \label{eq:TimeEquiv}
\end{equation}
Next, we take the derivative of $l'$,

\begin{equation}
    dl' = ct \cosh\left(\chi\right) d\chi + c \sinh(\chi) dt.
\end{equation}
Along a constant time-slice ($dt = 0$);

\begin{equation}
    dl' = ct \cosh\left(\chi\right) d\chi. \label{eq:dlConstTimeslice}
\end{equation}
Substitution the space relation of Equation~\ref{eq:FLRWMilneConversion},  along with Equtions \ref{eq:TimeEquiv} and \ref{eq:dlConstTimeslice} into our standard Lorentz boosted Minkowski metric yields

\begin{equation}
    ds^2 = -c^2 dt^2 + c^2 t^2 \left( d\chi^2 + \sinh^2(\chi) \right).
\end{equation}
This provides the form of the standard FLRW metric (Equation 1) with a curvature parameter $k = -1$, $R(t) = ct$, along a radial line of sight ($d\theta = d\phi = 0$).

\end{appendix}

\bibliographystyle{pasa-mnras}
\bibliography{bibliography.bib}

\end{document}